\documentclass[bm,aps,showpacs,amsfonts,amssymb]{revtex4}  

\usepackage{graphicx,bm}
\usepackage{natbib}

\usepackage{color}
\newcommand\beqa{\begin{eqnarray}}
\newcommand\eeqa{\end{eqnarray}}

\begin{document}

{~}

\title{
Non-linear effects for cylindrical gravitational two-soliton}
\vspace{2cm}
\author{Shinya Tomizawa${}^{1}$\footnote{tomizawasny@stf.teu.ac.jp} and Takashi Mishima${}^2$\footnote{tmishima@phys.ge.cst.nihon-u.ac.jp}}
\vspace{2cm}
\affiliation{
${}^1$ Department of Liberal Arts, Tokyo University of Technology, 5-23-22, Nishikamata, Otaku, Tokyo, 144-8535, Japan, \\
${}^2$ Laboratory of Physics, College of Science and Technology, Nihon University,
Narashinodai, Funabashi, Chiba 274-8501, Japan}

\begin{abstract} 
Using a cylindrical soliton solution to the four-dimensional vacuum Einstein equation, we study non-linear effects of gravitational waves such as Faraday rotation and time shift phenomenon. In the previous work, we analyzed the single-soliton solution constructed by the Pomeransky's improved inverse scattering method.  In this work, we construct a new two-soliton solution with complex conjugate poles, by which we can avoid light-cone singularities unavoidable in a single soliton case. In particular, we compute amplitudes of such non-linear gravitational waves and  time-dependence of the polarizations.  Furthermore, we consider the time shift phenomenon for soliton waves, which means that a wave packet can propagate at slower velocity than light.

\end{abstract}

\date{\today}
\maketitle

\section{Introduction}
A lot of gravitational solitons in general relativity, which describe gravitational solitonic waves propagating throughout spacetime, have been found within the framework of  so-called inverse scattering method~\cite{book exact solution,soliton book}. In particular, it has attracted a lot of relativists since it can generate black hole solutions in an axisymmetric and stationary case~\cite{review IIM,review ER},  in addition to exact solutions describing non-linear gravitational waves on various physical background. 
 The method was immediately generalized to the higher dimensional Einstein equations (see the references in \cite{soliton book}) but such a simple generalization to higher dimensions tends to lead to singular solutions.
However, $10$ years ago, Pomeransky~\cite{Pom} succeeded in modifying the original inverse scattering method~\cite{Belinsky-Zakharov} so that it can generate regular solutions even in higher-dimensions. 
Thanks to this, as for five dimensions, several black holes solutions were found~\cite{review IIM,review ER}. 

\medskip
A diagonal metric form of a cylindrically symmetric spacetime makes the vacuum Einstein equation to an extremely simple structure of a linear wave equation in a flat background. 
As a solution of such a equation, Einstein-Rosen wave, which can be interpreted as  superposition of cylindrical gravitational waves with a $+$ mode only, were obtained~\cite{Einstein-Rosen, book exact solution}. However, the existence of nontrivial non-diagonal components of a metric drastically changes the structure of the Einstein equation since it generally yields a $\times$ mode together with non-linearity. Piran {\it et al}.~\cite{Piran} numerically studied non-linear interaction of cylindrical gravitational waves of both polarization modes and showed that a $+$ mode converts to a $\times$ mode. This phenomenon is named  {\it gravitational Faraday effect} after the Faraday effect in electrodynamics. 
Tomimatsu~\cite{tomimatsu} studied the gravitational Faraday rotation for cylindrical gravitational solitons generated by the inverse scattering technique. 
Moreover, the interaction of gravitational soliton waves with a cosmic string was also studied in~\cite{Economou,Xanthopoulos1,Xanthopoulos2}. As one of new attempts to understand strong gravitational effects we have recently constructed a new cylindrically symmetric  single-soliton solution from Minkowski seed by the Pomeransky's inverse scattering method and clarified the behavior of the new solution including the effect similar to the gravitational Faraday rotation.

\medskip
In this paper, for further investigation we first construct more complicated gravitational two-soliton solutions with complex conjugate poles by the Pomeransky's method, which are considered as cylindrically symmetric gravitational waves with rich structure and can be used to study gravitational non-linear effects.
One of important and remarkable features of the solutions is that there are no null singularities which appear generally in the case of single-soliton solutions. 
We may therefore adopt the picture similar to the scattering theory that from the past null infinity one gravitational wave packet comes into `the interaction region' near the symmetric axis and after reflection leave the region for the future null infinity. 
The behavior of the wave packet can be analyzed by following the time sequential images and also by comparing the physical quantities measured in the past and future infinity. 
As an interesting example, the change of polarization of two independent modes will be treated with the both methods

\medskip
In the next section, we present the Kompaneets-Jordan-Ehlers form~\cite{Komaneets-Jordan-Ehlers} in the most general cylindrical symmetric spacetime and the useful  quantities (amplitudes and polarization angles) for analysis of non-linear cylindrical gravitational waves, which were first introduced by Piran {\it et al.}\cite{Piran} and Tomimatsu~\cite{tomimatsu}  . 
In Sec.~\ref{sec:solution}, using the inverse scattering method improved by the Pomeransky~\cite{Pom} , we will generate a two-soliton solution with complex conjugate poles from Minkowski spacetime. 
In Sec.\ref{sec:Analysis} we will analyze the obtained two-soliton solution by computing the amplitudes and polarization angles for ingoing and outgoing waves. In this section, in particular, by seeing time-dependence of polarizations, we study gravitational Faraday effect. Furthermore, we will mention the difference from the single-soliton solution in our previous paper, and moreover clarify the difference from the Tomimatsu's two solution~\cite{tomimatsu} generated by the Belinsky-Zakharov's procedure. 
 In Sec.~\ref{sec:discuss}, we will devote ourselves to the summary and discussion on our results.

\section{Formulas}\label{sec:formulas}
We assume that a four-dimensional spacetime admits cylindrical symmetry, namely,  that there are two commuting Killing vector fields, an axisymmetric Killing vector $\partial/\partial \phi$ and a spatially translational Killing vector $\partial/\partial z$, where the polar angle coordinate $\phi$ and the coordinate $z$ have the ranges $0\le \phi<2\pi$ and $-\infty<z<\infty$, respectively.
Under these symmetry assumptions, the most general metric can be described  by the Kompaneets-Jordan-Ehlers form: 
\begin{eqnarray}
ds^2=e^{2\psi}(dz+\omega d\phi)^2+\rho^2 e^{-2\psi}d\phi^2+e^{2(\gamma-\psi)}(d\rho^2-dt^2),
\end{eqnarray}
where the functions $\psi$, $\omega$ and $\gamma$ depend on the time coordinate $t$ and radial coordinate $\rho$ only. 
Following Ref.~\cite{Piran,tomimatsu}, we introduce the amplitudes
\begin{eqnarray}
&&A_+=2\psi_{,v},\label{eq:Ap}\\
&&B_+=2\psi_{,u},\label{eq:Bp}\\
&&A_\times=\frac{e^{2\psi}\omega_{,v}}{\rho},\label{eq:Ac}\\
&&B_\times=\frac{e^{2\psi}\omega_{,u}}{\rho},\label{eq:Bc}
\end{eqnarray}
where the advanced ingoing and outgoing null coordinates $u$ and $v$ are defined by $u=(t-\rho)/2$ and $v=(t+\rho)/2$, respectively.
The indices $+$ and $\times$ denote the quantities associated with the respective polarizations. 
Then, the vacuum Einstein equation can be written in terms of these quantities, actually, the non-linear differential equations for the functions $\psi$ and $\omega$ are replaced by
\begin{eqnarray}
&&A_{+,u}=\frac{A_+-B_+}{2\rho}+A_\times B_\times,\\
&&B_{+,v}=\frac{A_+-B_+}{2\rho}+A_\times B_\times,\\
&&A_{\times,u}=\frac{A_\times+B_\times}{2\rho}-A_+ B_\times,\\
&&B_{\times,v}=-\frac{A_\times+B_\times}{2\rho}+A_\times B_+,
\end{eqnarray}
and the function $\gamma$ is determined by
\begin{eqnarray}
&&\gamma_{,\rho}=\frac{\rho}{8}\left(A_+^2+B_+^2+A_\times^2+B_\times^2\right),\\
&&\gamma_{,t}=\frac{\rho}{8}\left(A_+^2-B_+^2+A_\times^2-B_\times^2\right).
\end{eqnarray}

The ingoing and outgoing amplitudes are defined by, respectively, 
\begin{eqnarray}
A=\sqrt{A_+^2+A_\times^2},\quad B=\sqrt{B_+^2+B_\times^2},\label{eq:amplitues}
\end{eqnarray}
and the polarization angles $\theta_A$ and $\theta_B$ for the respective wave amplitudes are given by, 
\begin{eqnarray}
\tan2\theta_A=\frac{A_\times}{A_+},\quad\tan2\theta_B=\frac{B_\times}{B_+}.\label{eq:pola}
\end{eqnarray}

\section{Two-soliton solution}\label{sec:solution}
In this work, as a seed, we consider Minkowski spacetime written in cylindrical coordinates whose $2\times 2$ part of the metric $g_0:=(g_{0ab})\ (a,b=z,\phi)$ and metric function $f_0:=e^{2(\gamma_0-\psi_0)}$ are given by, respectively, 
\begin{eqnarray}
g_0={\rm diag}\left(1,\rho^2\right), \quad f_0=1.
\end{eqnarray}
Following the inverse scattering method which Pomeransky improved~\cite{Pom}, we construct a two-soliton solution with complex conjugate poles.  
First remove trivial solitons with $(1,0)$ at $t=a_1$ and $t=a_2$ ($\bar a_2=a_1$) from the seed metric and then we obtain a metric
\begin{eqnarray}
&&g_0'={\rm diag}\left(\frac{|\mu_1|^4}{\rho^4},\rho^2\right)=(|w|^4,\rho^2),\label{eq:seed}
\end{eqnarray}
where $w:=\mu_1/\rho\ (\mu_1=\bar\mu_2=\sqrt{(t-a_1)^2-\rho^2}-(t-a_1))$. Note that $a_1$ is a complex parameter and the bar denotes complex conjugation. 
Next add back non-trivial solitons with $(1,a)$ and $(1,\bar a)$. 
Then we obtain two-soliton solution as
\begin{eqnarray}
&&g_{ab}=(g_0)_{ab}-\sum_{k,l=1}^2\frac{(g_0)_{ac}m_c^{(k)}(\Gamma^{-1})_{kl}m_d^{(l)}(g_0)_{db}}{\mu_k\mu_l},\\
&&f=f_0\frac{{\rm det} (\Gamma_{kl})}{{\rm det} (\Gamma_{kl}(c=0))},
\end{eqnarray}
where
\begin{eqnarray}
&&\Gamma_{kl}=\frac{m_a^{(k)}(g_0)_{ab}m_b^{(l)}}{-\rho^2+\mu_k\mu_l},\\
&&m_a^{(k)}=m_{0b}^{(k)}(\Psi^{-1}_{0}(\rho,t,\mu_k))_{ba}.
\end{eqnarray}
$\Psi_0(\rho,t,\lambda)$ is a generating matrix for the seed $g_0'$, which is given by
\begin{eqnarray}
 \Psi_0(\rho,t,\lambda)={\rm diag}\left(\frac{(\rho^2+2t\lambda+\lambda^2)^2}{(\tilde \mu_1-\lambda)^2(\tilde\mu_2-\lambda)^2},\rho^2+2t\lambda+\lambda^2\right),
\end{eqnarray}
where $\tilde\mu_k=\rho^2/\mu_k\ (k=1,2)$.

\medskip
We present the metric in the Kompaneets-Jordan-Ehlers form~\cite{Komaneets-Jordan-Ehlers}, which describes the most general cylindrically symmetric spacetime,  as
\begin{eqnarray}
ds^2=e^{2\psi}(dz+\omega d\phi)^2+\rho^2 e^{-2\psi}d\phi^2+e^{2(\gamma-\psi)}(d\rho^2-dt^2),
\end{eqnarray}
where the functions $\psi$, $\omega$ and $\gamma$ depend on the time coordinate $t$ and radial coordinate $\rho$ only and they are explicitly  given by
\begin{eqnarray}
&&e^{2\psi}=|w|^4\left(1-\frac{{\cal A}}{{\cal B}}\right),\\
&&\omega=-\frac{(|w|^2-1)^2}{\rho}\frac{{\cal C}}{{\cal B}-{\cal A}},\\
&&e^{2\gamma}=\frac{|w|^4({\cal B}-{\cal A})}{(w-\bar w)^2|w^2-1|^6(|w|^2-1)^6},
\end{eqnarray}
with
\begin{eqnarray}
{\cal A}&=&2\Re\left[\frac{(|w|^2-1)^4(\bar w^2-1)^4}{\bar w^2(w^2-1)}(X^2+c^2Y^2)\right]-2\frac{(|w|^2-1)^3|w^2-1|^4}{|w|^2}(|X|^2+|c|^2|Y|^2),\\
{\cal B}&=&\frac{1}{|w^2-1|^2}| X^2+c^2Y^2 |^2-\frac{1}{(|w|^2-1)^2}(|X|^2+|c|^2|Y|^2)^2,\\
{\cal C}&=&2\Re\left[\frac{\bar c(\bar w^2-1)^2}{\bar w(w^2-1)}(X^2+c^2Y^2)\right]-2\Re\left[\frac{\bar c(w^2-1)^2}{w(|w|^2-1)}\right](|X|^2+|c|^2|Y|^2),
\end{eqnarray}
\begin{eqnarray}
&&X=(w^2-1)^2(|w|^2-1)^2,\\
&&Y=\frac{|w|^2w}{\rho^2},\\
&&c=2aa_1.
\end{eqnarray}
Here, $\Re[\ ]$ denotes a real part of $[\ ]$.

\medskip
After using the time translational invariance of the system to rewrite the parameter $a_{1}$ as $i q$ ($i$ is $\sqrt{-1}$ and $q$ is a positive number), 
we can simplify the metric by introducing the following new coordinates $(x,y)$:
\begin{eqnarray}
t=qx y,\quad \rho=q\sqrt{(x^2+1)(y^2-1)}.
\end{eqnarray}
Note that 
\begin{eqnarray}
\mu_1=\bar \mu_2=q(x+i)(1-y).
\end{eqnarray}
Let us put $a=a_r+a_i i$ $(a_r,a_i \in {\bm R})$. 
In the coordinate system $(x,y)$, the metric can be written as
\begin{eqnarray}
ds^2&=&\frac{{\cal Y}}{\cal X}\left(dz+\frac{\cal Z}{\cal Y}d\phi\right)^2+\rho^2\frac{\cal X}{\cal Y}d\phi^2+\frac{\cal X}{4096q^4(x^2+y^2)^5}\left(-\frac{dx^2}{x^2+1}+\frac{dy^2}{y^2-1}\right),
\end{eqnarray}
where the metric functions $\cal X,\cal Y$ and $\cal Z$ are
\begin{eqnarray*}
\cal X&=&a_i^4 (y-1)^2 (y+1)^6+2 a_i^2 (y+1)^2 \left(a_r^2 (y-1)^2 (y+1)^4+64 q^2 \left(x^4 (y (9 y-8)+1)+2 x^2 (y (y+4)-3) y^2+y^6+y^4\right)\right)\\
&&-512 a_i a_r q^2 x (y+1)^2 \left(x^2-(y-2) y\right) \left(x^2 (2 y-1)+y^2\right)+a_r^4 (y-1)^2 (y+1)^6\\
&&+128 a_r^2 q^2 (y+1)^2 \left(2 x^6+x^4 ((8-3 y) y-1)+2 x^2 y^2 (2 (y-2) y+3)+y^6-y^4\right)+4096 q^4 \left(x^2+y^2\right)^4,\\
\cal Y&=&a_i^4 \left(y^2-1\right)^4+2a_i^2 \left(y^2-1\right) \left(a_r^2 \left(y^2-1\right)^3+64 q^2 \left(x^4 \left(9 y^2-1\right)+2 x^2 \left(y^2+3\right) y^2+y^6-y^4\right)\right)\\
&&-1024 a_i a_r q^2 x \left(x^2+1\right) y \left(y^2-1\right) (x-y) (x+y)+a_r^4 \left(y^2-1\right)^4\\
&&+128 a_r^2 q^2 \left(y^2-1\right) \left(2 x^6+x^4+\left(4 x^2+1\right) y^4-3 \left(x^2+2\right) x^2 y^2+y^6\right)+4096 q^4 \left(x^2+y^2\right)^4,\\
\cal Z&=&-32 q^2 (y+1) (a_i^3 x (y-1) (y+1)^3 \left(x^2 (1-3 y)+(y-3) y^2\right)+a_i^2 a_r (y-1) (y+1)^3 \left(x^4-3 x^2 (y-1) y-y^3\right)\\
&&+a_i x (y-1) \left(a_r^2 (y+1)^3 \left(x^2 (1-3 y)+(y-3) y^2\right)+64 q^2 \left(x^2+y^2\right)^3\right)\\
&&+a_r \left(a_r^2 (y-1) (y+1)^3 \left(x^4-3 x^2 (y-1) y-y^3\right)-64 q^2 \left(x^2+y\right) \left(x^2+y^2\right)^3\right).
\end{eqnarray*}

\section{Analysis}\label{sec:Analysis}
Let us investigate how the new gravitational solitonic waves propagate throughout spacetime from the several viewpoints. 
For convenience of explanation, we introduce the modulus $k$ and the angle $\theta$ of the complex  parameter $a$ defined by eq. (33), 
\begin{eqnarray*}
k=|a|,\quad \theta={\rm Arg}(a).
\end{eqnarray*}
In the following analysis, we only consider the case of $q=1$, because the parameter $q$ can be normalized by the scaling of coordinates. 

\begin{figure}[!h]
 \begin{center}
  \includegraphics[width=150mm]{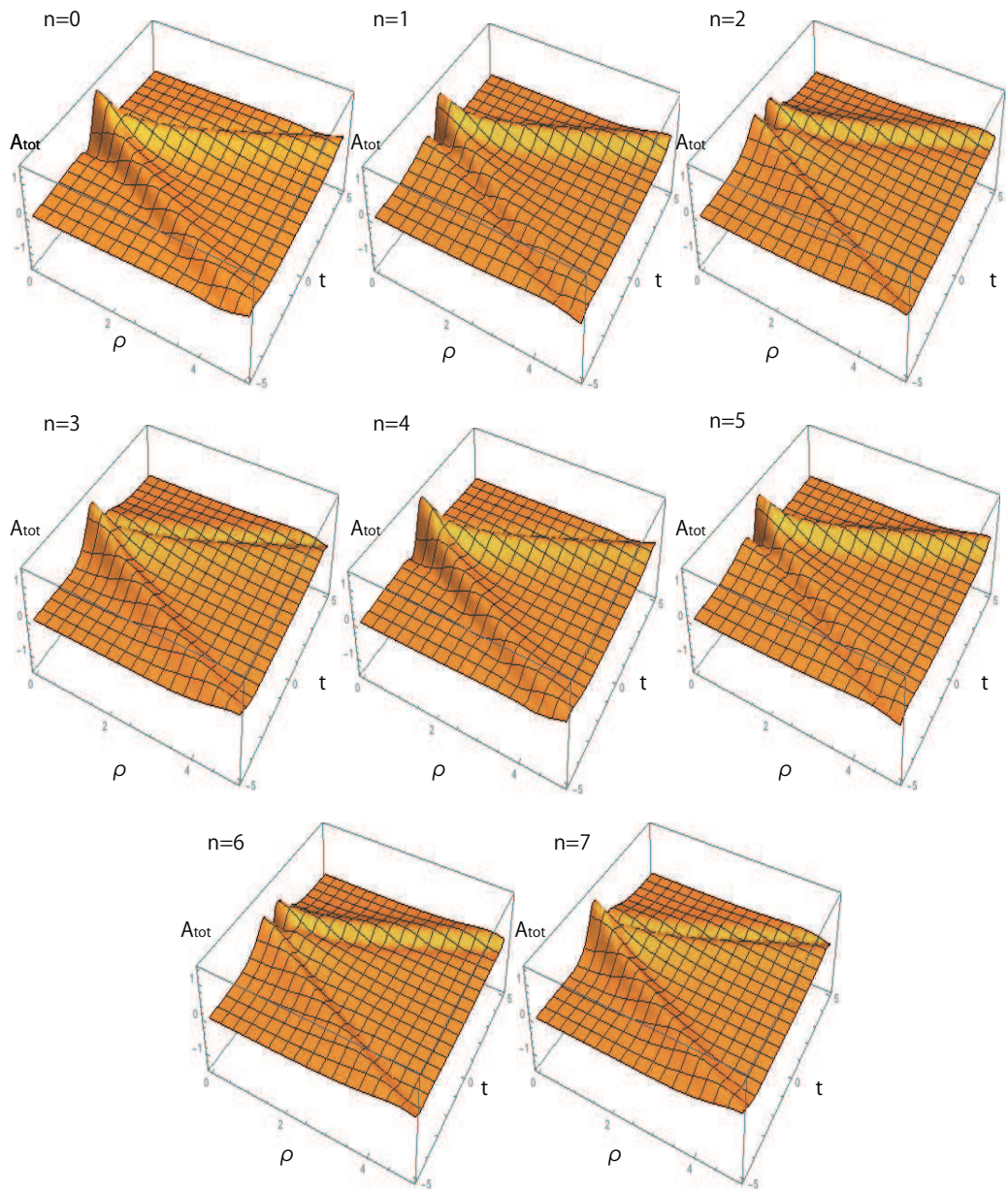}
 \end{center}
 \caption{The total amplitude $A_{\rm tot}=\sqrt{A^2+B^2}$ in a $(\rho,t)$-plane for $(k,\theta)=(2,n\pi/4)\ (n=0,1,\cdots,7)$.}
 \label{fig:k2theta0}
\end{figure}

\medskip
At the beginning we show the qualitative behavior of the waves near the axis briefly. 
The figures $n=0,\cdots, 7$ in FIG.\ref{fig:k2theta0} display the various behaviors of total amplitude ($A_{\rm tot}=\sqrt{A^2+B^2}$) from the spacetime viewpoint, which are arranged from left to right in each figure, respectively. 
The figures are plotted in the range of $-5\le t \le 5$ and $0\le \rho\le 5$ under the parameter-setting $q=1,$ $k=2$ and $\theta=n \pi/4\ (n=0\sim 7)$. 
Some of the behaviors are also displayed in FIG.\ref{fig:1timedependentk2theta0}  by superimposing the instantaneous graphs that correspond to $t=0$ and $\ t=\pm n,\ (n=1\sim 3)$, respectively. 
From the behaviors of the amplitudes we may consider the gravitational solitonic waves to be regular wave packets, which first come into the region near the symmetric axis from past null infinity, and leave the axis after reflection for the future null infinity. 
On first inspection the various complex behaviors seem to be generated by a certain kind of non-linear effect near the axis.
For example, shown in the case of $n=1$ in FIG.\ref{fig:k2theta0} and  FIG.\ref{fig:1timedependentk2theta0}, at the angle $\theta=\pi/4$ the initial two wave packets coalesce into one packet, while at the angle $\theta=3\pi/4$ the initial single wave packet splits in two.  
These remarkable phenomena seem to occur very near the axis, so that we may say that the neighborhood of the axis is the region where non-linear effects become highly strong. 

\medskip
In the subsequent subsections, we will see the detailed behavior of the waves propagating near the  boundaries of spacetime, particularly focusing on its non-linear effect. 

\subsection{Initial amplitudes}
\begin{figure}[!h]
 \begin{center}
  \includegraphics[width=140mm]{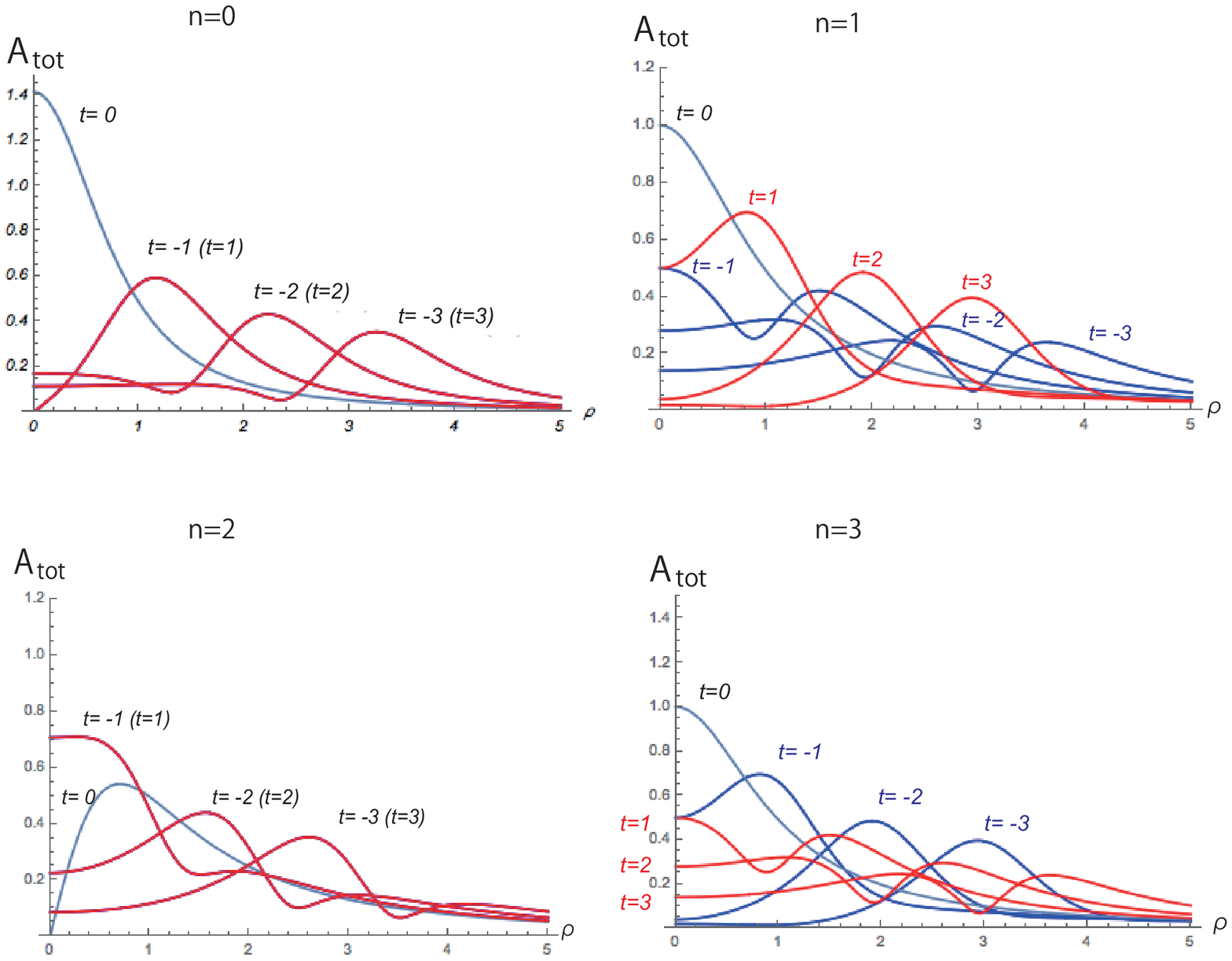}
 \end{center}
 \caption{Time-dependence of total amplitude $A_{\rm tot}$ for $(k,\theta,q)=(2,n\pi/4,1)\ (n=0,1,2,3)$: The blue-colored and red-colored graphs denote incident waves and reflectional waves, respectively, for $t=,\pm 1,\pm 2,\pm 3$. The violet-colored graphs in the case  of $n=0,2$ mean that incident and reflectional waves entirely overlap.}
 \label{fig:1timedependentk2theta0}
\end{figure}

Let us see the initial amplitudes for the ingoing and outgoing waves. 
At the initial time $t=0$, we have the ingoing and outgoing wave amplitudes
\begin{eqnarray}
A=\frac{4q}{\sqrt{\rho^2+q^2}}\sqrt{\frac{N+128a_ia_rq^3(\rho^2+q^2)\rho}{D}},\\
B=\frac{4q}{\sqrt{\rho^2+q^2}}\sqrt{\frac{N-128a_ia_rq^3(\rho^2+q^2)\rho}{D}},
\end{eqnarray}
where
\begin{eqnarray*}
&&N:=(|a|^4+64a_i^2q^2)\rho^6+64q^4(|a|^2+a_i^2)\rho^4+64q^6(|a|^2+a_r^2)\rho^2+64q^8a_r^2,\\
&&D:=(|a|^2+64q^2)^2\rho^8+256q^4(|a|^2+a_r^2+64q^2)\rho^6+128q^6(|a|^2+4a_r^2A_{\rm tot}+192q^2)\rho^4+256q^8(a_r^2+64q^2)\rho^2+4096q^{12}.
\end{eqnarray*}
As is shown in FIG.\ref{fig:1timedependentk2theta0}, the disturbances for the  total  amplitude $A_{\rm tot}$, which is proportional to  the {\it C}-energy density, is localized in the neighborhood of the axis.

\subsection{Asymptotic behaviors}
\subsubsection{Timelike infinity}
Next we consider the asymptotic behaviors of the waves at late time $t\to\infty$.
At $t\to\infty$, the metric behaves as
\begin{eqnarray}
ds^2\simeq \left(1-\frac{a_r^2}{4t^2}\right)\left[dz+a_r\left(1+\frac{\rho^2}{4t^2}\right)d\phi \right]^2+\rho^2\left(1+\frac{a_r^2}{4t^2}\right)d\phi^2+\left(1+\frac{a_r^2}{4t^2}\right)(-dt^2+d\rho^2).
\end{eqnarray}
Here let us introduce the new coordinate $\tilde z:=z+a_r\phi$ so that $\partial/\partial \tilde z$ is also a translationally symmetric Killing vector, and then we can show that this asymptotic metric is that of the Minkowski spacetime. 
Therefore, both ingoing and outgoing waves vanish at late time. 
We have the asymptotic amplitudes
\begin{eqnarray}
A\simeq B\simeq \frac{a_r}{2t^2}+{\cal O}(t^{-3}).
\end{eqnarray}
In particular, for $a_r=0$, the amplitudes  behave in a different way as
\begin{eqnarray}
A\simeq B\simeq \frac{a_iq}{t^3}+{\cal O}(t^{-4}). 
\end{eqnarray}
In this region, the polarization angles $\theta_A$ and $\theta_B$ for outgoing and ingoing waves behaves as
\begin{eqnarray}
\tan\theta_A\simeq\tan\theta_B\simeq 1.
\end{eqnarray}
This shows that regardless of the values of the parameters, the $\times$ mode  becomes dominate at late time.

\subsubsection{Spacelike infinity}
Let us consider gravitational waves near spacelike infinity $\rho\to\infty$. 
In the limit of $\rho\to\infty$, we have the following asymptotic metric form
\begin{eqnarray}
ds^2\simeq \left(1-\frac{4|a|^2q}{(|a|^2+64q^2)\rho}\right)\left[dz+\frac{32a_rq^2}{|q|^2+64q^2}d\phi\right]^2+\rho^2 \left(1+\frac{4|a|^2q}{(|a|^2+64q^2)\rho}\right)d\phi^2+\frac{(|a|^2+64q^2)^2}{4096q^4}(-dt^2+d\rho^2).
\end{eqnarray}
If we use the new coordinate 
\begin{eqnarray}
\tilde z&=&z+\frac{32a_rq^2}{(|a|^2+64q^2)}\phi,
\end{eqnarray}
we immediately find that this spacetime is Minkowski spacetime with the deficit angle 
\begin{eqnarray}
D=2\pi\frac{|a|^2}{|a|^2+64q^2}.
\end{eqnarray}
The wave amplitudes behave as
\begin{eqnarray}
A\simeq B\simeq \frac{\sqrt{|a|^4+64a_i^2q^2}}{(|a|^2+64q^2)\rho^2}+{\cal O}(\rho^{-3}).
\end{eqnarray}
For $a_i\not=0$, the polarization angles approach the values
\begin{eqnarray}
\tan\theta_A\simeq -\tan\theta_B\simeq \frac{|a|^2-\sqrt{|a|^4+16a_i^2q^2}}{16a_iq},
\end{eqnarray}
and for $a_i=0$, they behave as
\begin{eqnarray}
\tan\theta_A\simeq \tan\theta_B \simeq 0.
 \end{eqnarray}

\subsubsection{Axis}
Furthermore, let us see the behaviors of the waves on the axis of symmetry $\rho=0$.
Near the axis, the metric behaves as
\begin{eqnarray}
ds^2&\simeq& \frac{4(q^2+t^2)^2}{4(t^2+q^2)^2+(ta_r-qa_i)^2}\left(dz+a_rd\phi\right)^2+ \frac{4(t^2+q^2)^2+(ta_r-qa_i)^2}{4(q^2+t^2)^2}\rho^2d\phi^2\nonumber\\
&&+\frac{4(t^2+q^2)^2+(ta_r-qa_i)^2}{4(q^2+t^2)^2}(-dt^2+d\rho^2).
\end{eqnarray}
Using the coordinate $\tilde z=z+a_r \phi$, we have the metric
\begin{eqnarray}
ds^2\simeq \frac{4(q^2+t^2)^2}{4(t^2+q^2)^2+(ta_r-qa_i)^2}d\tilde z^2+ \frac{4(t^2+q^2)^2+(ta_r-qa_i)^2}{4(q^2+t^2)^2}\rho^2d\phi^2+\frac{4(t^2+q^2)^2+(ta_r-qa_i)^2}{4(q^2+t^2)^2}(-dt^2+d\rho^2).
\end{eqnarray}
It is straightforward to show that the ratio of length of the circumference to the radial distance on the axis is
\begin{eqnarray}
\lim_{\rho\to0}\frac{\int^{2\pi}_0\sqrt{g_{\phi\phi}}d\phi}{\int^\rho_0\sqrt{g_{\rho\rho}}d\rho}
=2\pi.
\end{eqnarray}
This equation means that no deficit angle is present on the axis, which is in contrast to spacelike infinity. 
In the limit of $\rho\to 0$, the wave amplitudes behave as
\begin{eqnarray}
A\simeq B\simeq \frac{|a_rt^2-2a_iqt-a_rq^2|\sqrt{16(t^2+q^2)^2(a_iq-a_rt)^2+\{(a_rt-a_iq)^2-4(t^2+q^2)^2\}^2}}{2(t^2+q^2)^2[(a_rt-a_iq)^2+4(q^2+t^2)^2]}.
\end{eqnarray}
The polarizations are
\begin{eqnarray}
\tan2\theta_A\simeq \tan2\theta_B\simeq -\frac{(2t^2+2q^2-a_rt+a_iq)(2t^2+2q^2+a_rt-a_iq)}{4(t^2+q^2)(a_iq-a_rt)}.
\end{eqnarray}

\subsubsection{Null infinity}
At null infinity $v\to\infty$, the wave amplitudes behaves as 
\begin{eqnarray}
&&A\simeq 2p_1\sqrt{\frac{N_1}{D_1v^3}},\\
&&B\simeq \frac{8p_1^3}{p_1^4+q^2}\sqrt{\frac{N_2}{D_2v}},
\end{eqnarray}
where
\begin{eqnarray}
N_1&=&(|a|^4q^4+64a_i^2q^6)-128a_ra_iqp_1^2(q^2+p_1^4)^2+64(|a|^2+a_i^2)q^4p_1^4+64(|a|^2+a_r^2)q^2p_1^8+64a_r^2p_1^{12},\\
D_1&=&q^4(|a|^2+64q^2)^2+256q^4(|a|^2+a_r^2+64q^2)p_1^4+1024a_ra_iqp_1^6(q^2-p_1^4)+384q^2(3|a|^2-4a_r^2+64q^2)p_1^8\nonumber\\
&&+256(a_r^2+64q^2)p_1^{12}+4096p_1^{16},\\
N_2&=&|a|^4q^4+64a_i^2q^6-384a_ra_ip_1^2q(q^4+p_1^8)+192q^2p_1^4(|a|^2-5a_i^2)(q^2-p_1^4)+1280a_ra_iq^3p_1^6+64a_r^2p_1^{12},\\
D_2&=&q^4(|a|^2+64q^2)^2+256q^4(|a|^2+a_r^2+64q^2)p_1^4+1024a_ra_iqp_1^6(q^2-p_1^4)+384q^2(3|a|^2-4a_r^2+64q^2)p_1^8\nonumber\\
&&+256(a_r^2+64q^2)p_1^{12}+4096p_1^{16},
\end{eqnarray}
and $p_1=\sqrt{2u+\sqrt{4u^2+q^2}}$.

\subsection{Faraday effect}\label{sec:faraday}
\begin{figure}[!h]
 \begin{center}
  \includegraphics[width=80mm]{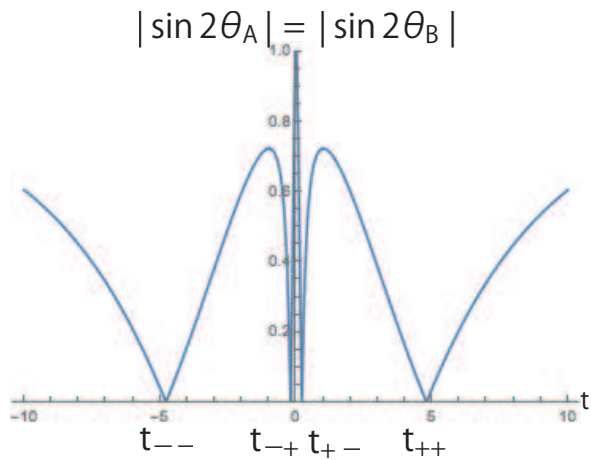}
 \end{center}
 \caption{The behaviors of  $|\sin2\theta_A |$ and $|\sin2\theta_B |$ for $(a_r,a_i,q)=(10,0,1)$ on the axis}
 \label{fig:1axisar10ai0}
\end{figure}

In this subsection, using the two-soliton solution, we study gravitational Faraday effect, which is a phenomenon that an outgoing (ingoing) wave amplitude corresponding to the $+$ mode converts to an outgoing (ingoing) wave amplitude corresponding to the $\times$ mode due to the interaction with an ingoing (outgoing) wave with the $\times$ mode.  
Let us see how the $+$ modes convert to the $\times$ modes while the corresponding gravitational waves are propagating along null rays from the axis $\rho=0$ to null infinity $v\to\infty$. It is the most interesting to analyze it in the cases of $a_r^2-16q^2-8a_iq>0$ and $a_r^2-16q^2+8a_iq>0$ since both the polarization angles $\theta_A$ and $\theta_B$ completely vanish on the axis four times,  i.e., at $t=t_{\pm\pm}$, where $t_{\pm\pm}$ are defined by, respectively, 
\begin{eqnarray}
t_{+\pm}&=&\frac{a_r \pm\sqrt{a_r^2-16q^2- 8 a_i q}}{4},\\
t_{-\pm}&=&\frac{-a_r \pm\sqrt{a_r^2-16q^2+ 8 a_i q}}{4}.
\end{eqnarray}
As is shown in FIG.\ref{fig:1axisar10ai0}, $\theta_A$ and $\theta_B$ have time-dependence on the axis $\rho=0$ and no $\times$ mode is present there just only at that moments.

\begin{figure}[!h]
\begin{center}
  \includegraphics[width=120mm]{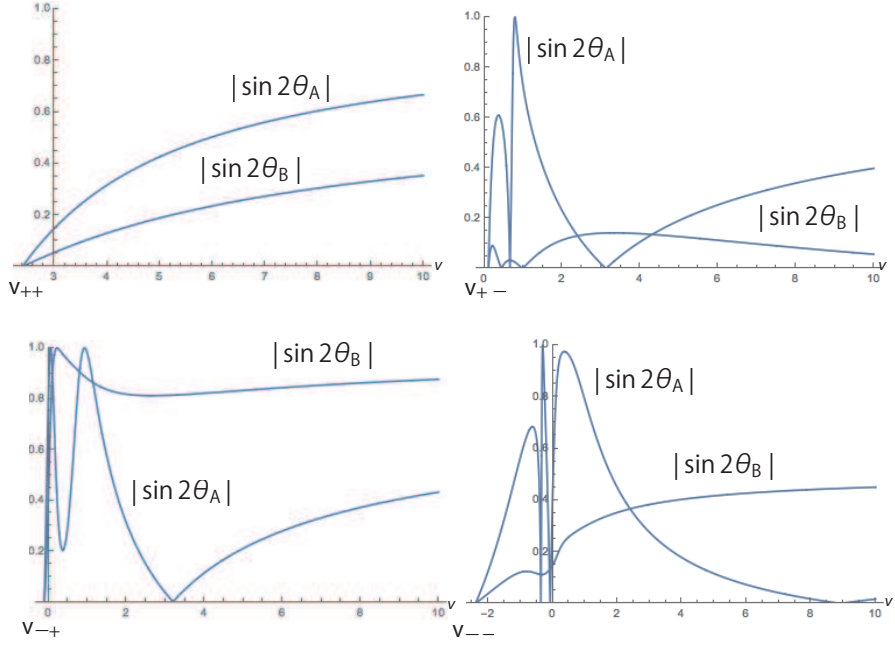}
 \end{center}
\caption{$|\sin 2\theta_A|$ and $|\sin 2\theta_B|$ along the outgoing null rays $u_{++}$, $u_{+-}$, $u_{-+}$,$u_{--}$  for $(a_r,a_i,q)=(10,0,1)$}
 \label{fig:1thetaAB-null-ar10ai0}
\end{figure}

\medskip
The upper-left graph in FIG.\ref{fig:1thetaAB-null-ar10ai0} shows that the pure $+$ mode wave passing $(t,\rho)=(t_{++},0)$ partially converts to the $\times$ mode wave and that its conversion comes to stop asymptotically at $v\to\infty$. The ratio of the $\times$ mode to the $+$ mode monotonously approaches to a certain constant value. 
In this process, a complete conversion of $+$ to $\times$, or of $\times$ to $+$ does not occur. 
The upper-right graph in FIG.\ref{fig:1thetaAB-null-ar10ai0} illustrates that the pure $+$ gravitational wave passing $(t,\rho)=(t_{+-},0)$ converts a little to the $\times$ mode and soon converts to the pure $+$ mode. After that, a little conversion occurs and again becomes the pure $+$ mode twice and $\theta_B$ approaches to a non-zero value. Finally, its conversion comes to stop asymptotically at $v\to\infty$. The ratio of the $\times$ mode to the $+$ mode becomes a certain constant value. 
The lower-left graph in FIG.\ref{fig:1thetaAB-null-ar10ai0} shows the pure $+$ gravitational wave passing $(t,\rho)=(t_{-+},0)$ completely converts to the $\times$ mode and after that partially converts to the $+$ mode. At $v\to \infty$, the $\times$ mode becomes dominate.
The lower-right graph in FIG.\ref{fig:1thetaAB-null-ar10ai0} illustrates that the pure $+$ gravitational wave passing $(t,\rho)=(t_{--},0)$ partially converts to the $\times$ mode.
At $v\to \infty$, $\theta_B$ asymptotically approaches to a certain constant.

\begin{figure}[!h]
\begin{center}
  \includegraphics[width=80mm]{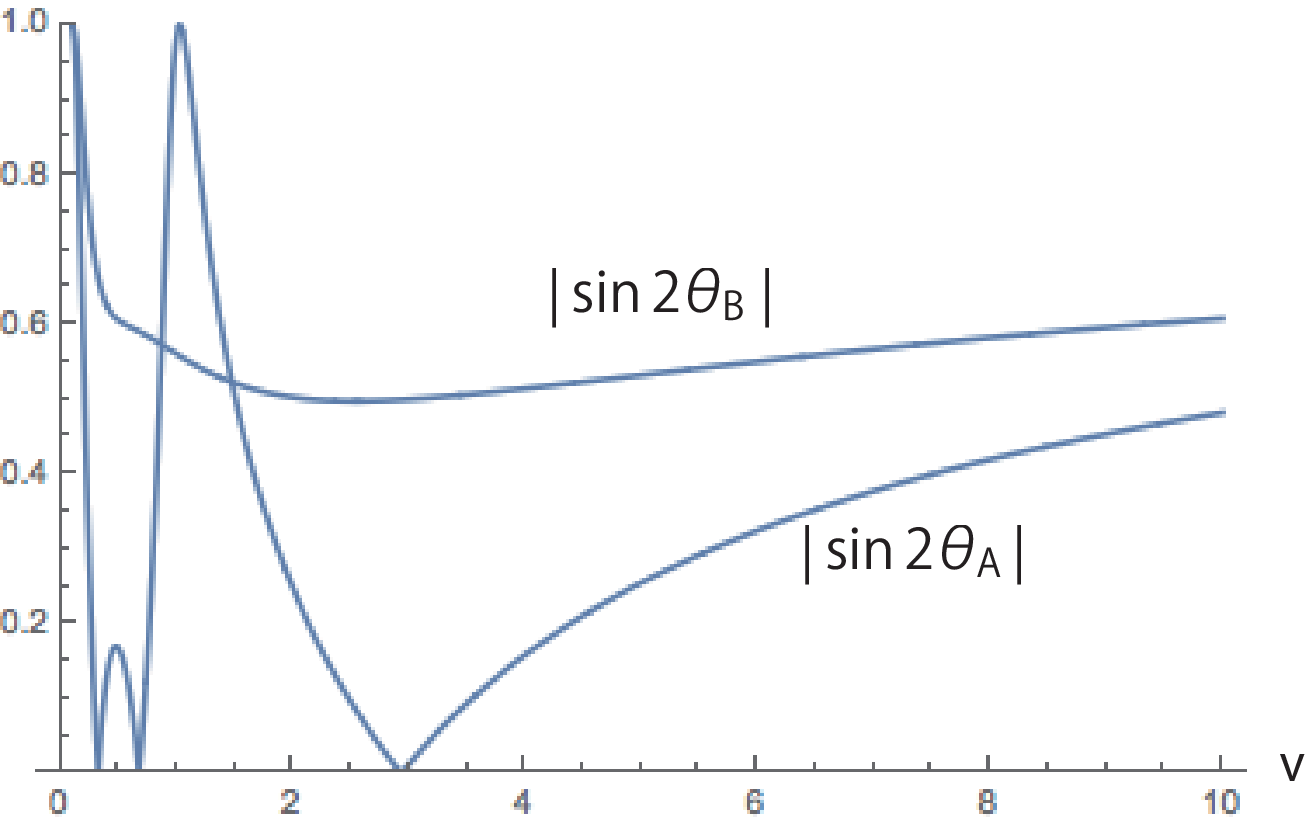}
 \end{center}
\caption{$|\sin 2\theta_A|$ and $|\sin 2\theta_B|$ along the outgoing null ray $u_{\times}$ for $(a_r,a_i,q)=(10,2,1)$}
 \label{fig:1thetaABar10ai2}
\end{figure}

\medskip
 
Finally, let us consider  the gravitational wave passing $(t,\rho)=(t_\times,0)$, where $t_\times=a_iq/a_r$, when no $+$ mode is present at the axis. 
As shown in FIG.\ref{fig:1thetaABar10ai2}, the pure $\times$ mode partially converts to the $+$ mode but  $\times$ mode again increases, and at $v\to \infty$, the value of $\theta_B$ becomes a constant.

\subsection{Time shift}
\begin{figure}[!h]
\begin{center}
  \includegraphics[width=120mm]{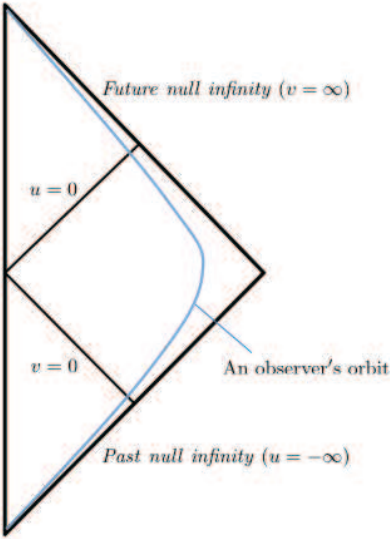}
 \end{center}
\caption{An orbit of the massless particle which is propagating in an incoming radial direction along $v=0$ and after reflection at the axis $\rho=0$ is propagating in an outgoing direction  along $u=0$. }
 \label{fig:time-shift}
\end{figure}

Let us investigate a time shift phenomenon as the non-linear effect, which means that a wave packet propagates at slower speed than light velocity. 
Basically following the analysis in Ref.\cite{ Dagotto}, where how to measure a time shift for gravitation solitons was proposed,  we numerically analyze the asymptotic behavior of the wave packets at future null infinity $v\to \infty$ and past null infinity $u\to-\infty$. In principle, we can find a time shift of the wave amplitudes by comparing its arrival time at future null infinity with that of a massless test particle starting off past null infinity at the same time. As is illustrated in FIG.\ref{fig:time-shift},  an incoming massless particle propagating along $v=0$ from past null infinity $u=-\infty$ arrives at an axis $\rho=0$ and after reflection, in turns, it propagates along $u=0$ toward future null infinity $v=\infty$.  Let us see how slow a wave packet which has a peak near $u=0$ and $v=0$ is, compared with the massless particle. 

\begin{figure}[!h]
\begin{center}
  \includegraphics[width=120mm]{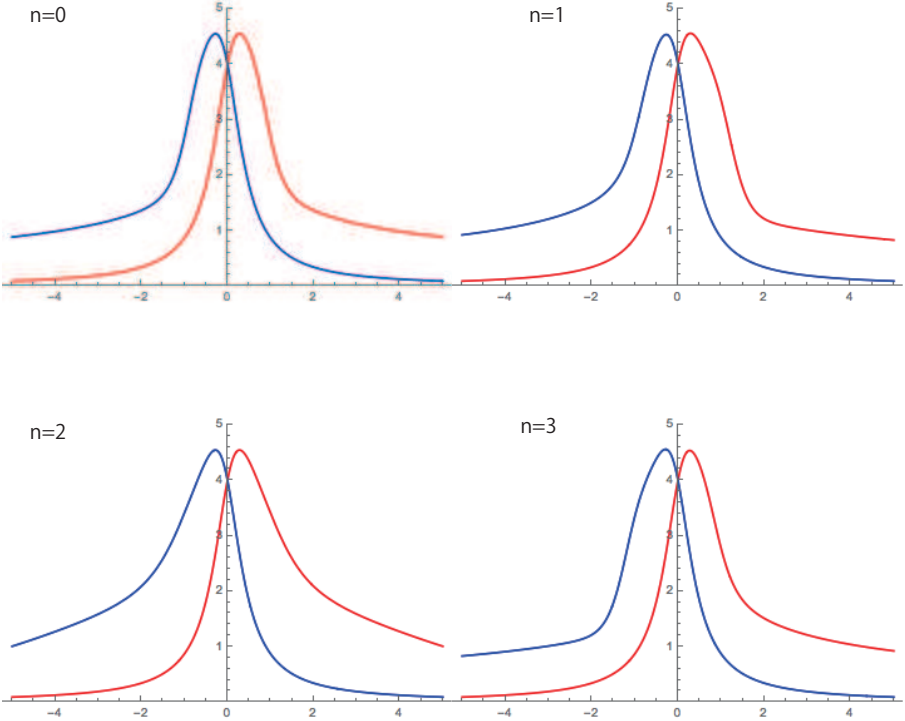}
 \end{center}
\caption{Time shift for $(k,\theta,q)=(1000,n\pi/4,1)$ $(n=0,1,2,3)$. The blue-colored and red-colored graphs show the incident waves at past null infinity and reflectional waves at future null infinity, respectively. }
 \label{fig:1timeshiftk1000}
\end{figure}

\medskip
In FIG.\ref{fig:1timeshiftk1000}, the blue-colored graphs denote the amplitudes for ingoing waves near past null infinity $\lim_{u\to-\infty}A\sqrt{-u}$, and 
the red-colored graphs denote the amplitudes for outgoing waves near future null infinity $\lim_{v\to\infty}B\sqrt{v}$ for $(k,\theta)=(2,n\pi/4)$   ($n=0,1,2,3$), 
where the amplitudes are multiplied by $(-u)^{\frac{1}{2}}$ and $v^{\frac{1}{2}}$ because of apparent decay at null infinity. We can interpret these results as follows. 
Let us consider an incoming massless test particle starting from past null infinity and propagating along a null geodesic $v=0$. The particle is reflected at an axis $\rho=0$ and then it propagates to future null infinity along a null geodesic $u=0$. 
An observer at past null infinity sees an ingoing wave packet earlier than an incoming radial photon, while at future null infinity he sees the outgoing wave packet after the outgoing photon. This shows a time shift phenomenon. 
\medskip
Moreover, we would like to comment that for the very large values, regardless of the difference between $\theta$'s values, the wave packet always propagate slower than a massless test particle, as is seen in FIG.~\ref{fig:1timeshiftk1000}. It can neither collide nor split in the process.

\subsection{Collision, coalescence and split of solitons}
\begin{figure}[!h]
\begin{center}
  \includegraphics[width=150mm]{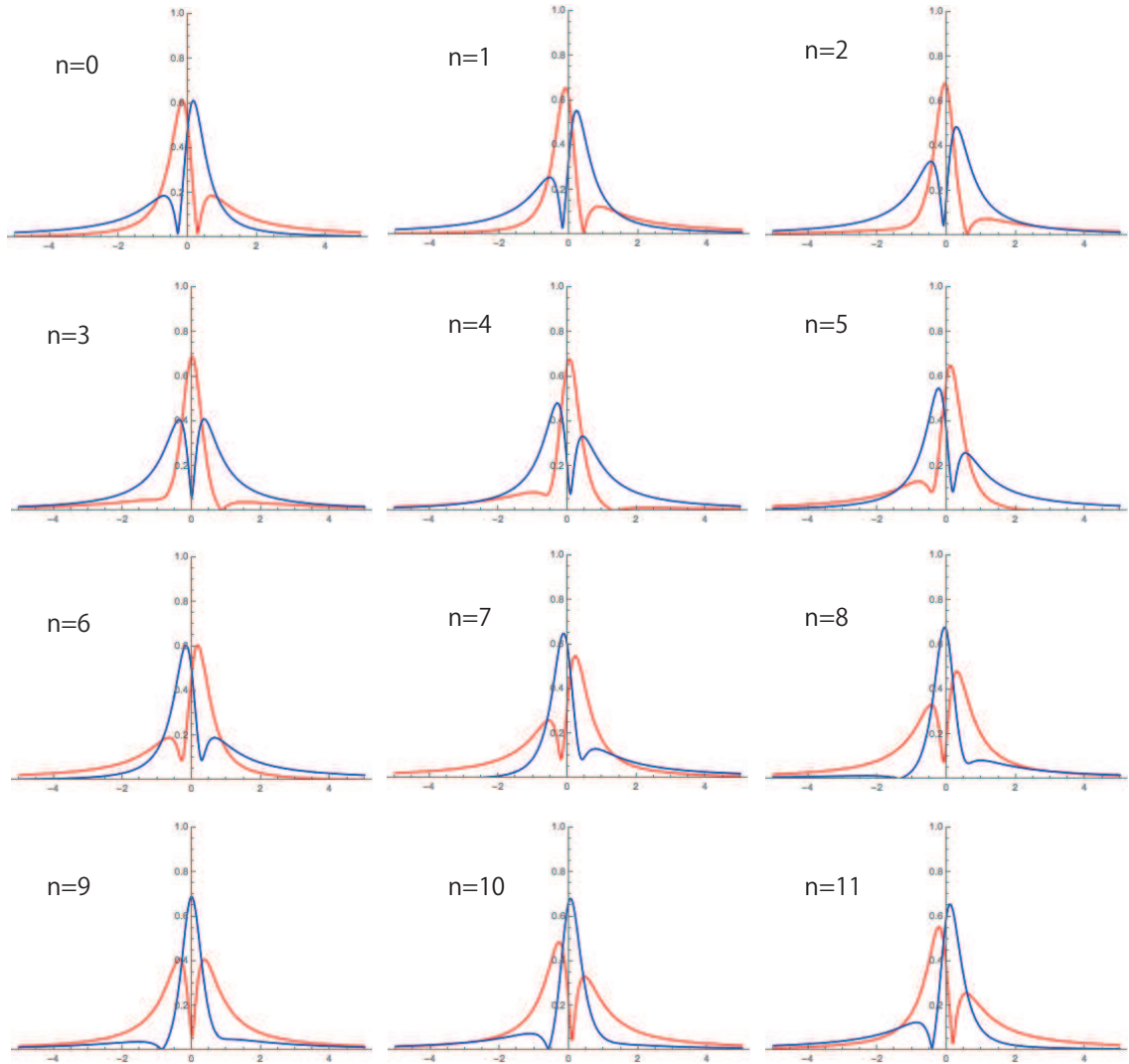}
 \end{center}
\caption{Amplitudes of the ingoing and outgoing waves for $(k,\theta,q)=(2,n\pi/12,1)$ $(n=0,1,\cdots,11)$. The blue-colored and red-colored graphs show the incident waves at past null infinity and reflectional waves at future null infinity, respectively. }
 \label{fig:1timeshiftk2}
\end{figure}

Besides the time shift phenomena, when $k\approx |q|$, the ingoing and outgoing waves take various shapes depending on the phase, as is seen in FIG.\ref{fig:1timeshiftk2}. 
As is seen in these graphs, at least, either of ingoing and outgoing waves can have two peaks. 

\medskip
For $n=0\ (\theta=0)$, there are two ingoing wave packets, one with a small peak and one with a large peak near past null infinity, and two outgoing wave packets, one with a small peak and one with a large peak near future null infinity. 
This obviously shows that two gravitational solitons collide, which occurs near the axis $\rho\simeq 0$, and then the larger one of two solitons overtakes the smaller one.
 For $n=6\ (\theta=\pi/2)$,  conversely, the smaller one collide with the larger one and then overtakes. 
 
 \medskip
For $n=3$ ($\theta=\pi/4$) , there are two ingoing solitons at past null infinity but a single outgoing soliton at near future null infinity.
This shows that two solitons coalesce (in reflection at the axis).
In contrast, as seen for $n=9$ ($\theta=3\pi/4$), a single wave packet splits into two wave packets. 
Such  phenomena do not happen for other solitons, such as solitons of KdV equation.

\section{Summary and Discussion}\label{sec:discuss}
In this paper, using the Pomeransky's inverse scattering method for a cylindrically symmetric spacetime and starting from the Minkowski seed, we have obtained the two-soliton solution which has two complex conjugate poles to the vacuum Einstein equation with cylindrical symmetry. 
As the one-soliton solution with a real pole (in our previous work~\cite{tomizawa-mishima}), it has been numerically shown that the two-soliton solution (presented in this work) describes 
a gravitational wave packet with two polarizations that comes from past null infinity, is reflected at the axis, and returns to future null infinity. 
The one-soliton solution in~\cite{tomizawa-mishima} describes a shock wave pulse with infinite amplitude propagating at light velocity, which yields null singularities, but the two-soliton solution is entirely free from such a singular behavior.  This fact itself should not be surprising because in the previous work~\cite{tomimatsu}, a two-soliton solution with complex conjugate poles and without any singularities has been constructed.  

\medskip
In this work, using the two-soliton solution, we have studied non-linear effect of cylindrically symmetric gravitational waves, focusing particularly on (i) gravitational Faraday effect, (ii) time shift phenomenon and (iii) collision process of two solitons:

\medskip
(i)  The polarization angles $\theta_A$ and $\theta_B$ of gravitational waves on the axis have time-dependence. 
In particular, if $a_r^2-16q^2-8a_iq>0$ or $a_r^2-16q^2+8a_iq>0$, at  the times $t=t_{\pm\pm}$ the $\times$ mode completely vanishes on the axis $\rho=0$ and the $+$ mode only is present there. 
In this case, we have studied how the pure $+$ mode on the axis converts to the $\times $ mode while it is propagating along the null rays $u=u_{\pm\pm}$. 
It is shown that in any cases, the polarization angles asymptotically a certain non-zero constant, which means that both modes are present at future null infinity.

\medskip
(ii) The time shift we say here is a phenomenon that a wave packet of a gravitational wave propagates at slower velocity than light. 
This is slightly different from the context used in the field of usual soliton theories, where this term (which is also called phase shift) is used to mean that when two solitonic waves collide,  each position shifts as compared with when it propagates alone.
In the case of cylindrical gravitational solitonic waves, it is evident that this phenomenon is due to self-interaction of a gravitational wave when an ingoing cylindrical wave reflects rather than to the interaction by a collision of two solitonic waves since in a region far from the axis, this gravitational soliton seems to propagate at light velocity.   

\medskip
 (iii) For the two-soliton solution in this paper, we have clarified that two gravitational solitons can coalesce into a single soliton and also a single soliton can split into two by non-linear effect of gravitation waves.  Such phenomena cannot be seen for solitons of other integrable equations such as solitons of KdV equation. For the KdV equation, when two solitons traveling in the same direction collide (the amplitude is not simply addition of the two individual solitons), each one soon separates from each other and then asymptotically comes to approach the same shape of a wave pulse as before the collision.


\begin{thebibliography}{99}
\bibitem{book exact solution}
H.~Stephani, D.~Kramer, M.~MacCallum, C.~Hoenselaers and E.~Herlt, {\it Exact solutions of Einstein's Field Equations, 2nd ed.} (Cambridge University Press, Cambridge, 2003).
\bibitem{soliton book}
V.~A.~Belinski and E.~Verdaguer, {\it Gravitational Solitons}, (Cambridge University Press, Cambridge, England, 2001).
\bibitem{review IIM}
H. Iguchi, K. Izumi and T. Mishima, Prog. Theor. Phys. Suppl. {\bf 189},  93-125 (2011).
\bibitem{review ER} 
R. Emparan and H. S. Reall, Living Rev. Rel. {\bf 11}, 6 (2008).
\bibitem{Pom}
A. A. Pomeransky, Phys. Rev. D {\bf 73},  044004 (2006).
\bibitem{Belinsky-Zakharov}
V. A. Belinskii and E. E. Zakharov, Sov. Phys. JETP, {\bf 49}, 985 (1979).
\bibitem{Einstein-Rosen}
A. Einstein and N. Rosen, J. Franklin Inst. {\bf 223}, 43 (1937).
\bibitem{Piran}
T. Piran, P. N. Safier and R. F. Stark,  Phys. Rev. D {\bf 32}, 3101 (1985).
\bibitem{tomimatsu}
A. Tomimatsu, Gen. Rel. Grav. {\bf 21}, 613 (1989).
\bibitem{Economou}
A. Economou and D. Tsoubelis, Phys. Rev. D {\bf 38}, 498 (1988).
\bibitem{Xanthopoulos1}
B. C. Xanthopoulos, Phys. Lett. B {\bf 178}, 163 (1986).
\bibitem{Xanthopoulos2}
B. C. Xanthopoulos, Phys. Rev. D {\bf 34}, 3608 (1986).
\bibitem{Komaneets-Jordan-Ehlers}
P. Jordan, J. Ehlers, and W. Kundt, Abh. Akad. Wiss. Mainz.
Math. Naturwiss. {\bf Kl}. 2 (1960);
A. S. Kompaneets, Zh. Eksp. Teor. Fiz. {\bf 34}, 953 (1958) [Sov.
Phys. JETP {\bf 7}, 659 (1958)].

\bibitem{tomizawa-mishima}
S. Tomizawa and T. Mishima, Phys. Rev. D {\bf 90}, 044036 (2014).



\bibitem{Rosen}
N. Rosen, Bull. Res. Counc. Isr. {\bf 3}, 328 (1953).
\bibitem{Dagotto}
A. D Dagotto, R. J Gleiser and C. O Nicasio, Class. Quant. Gravi {\bf 8}, 1185 (1991).

\end{thebibliography}
\end{document}